\documentclass[apj]{emulateapj}
\usepackage{apjfonts}

\newcommand{\suz}{{\it Suzaku}}
\newcommand{\swift}{{\it Swift}}
\newcommand{\chandra}{{\it Chandra}}
\newcommand{\integral}{{\it INTEGRAL}}

\newcommand{\fermi}{{\it Fermi}-LAT}
\newcommand{\wise}{{\it WISE}}
\newcommand{\uki}{{\it UKIDSS}}
\newcommand{\hess}{HESS~J1943+213}
\newcommand{\hs}{1ES~0229+200}
\newcommand{\newhs}{1ES~0347$-$121}

\slugcomment{Accepted by ApJ on 2014 April 11}

\shorttitle{\fermi\ and \suz\ observations of \newhs\ and \hess}
\shortauthors{Tanaka et al.}

\begin{document}

\title{Extreme Blazars Studied with {\it Fermi}-LAT and {\it Suzaku}:\\ 1ES~0347$-$121 and Blazar Candidate HESS~J1943+213 }

\author{Y.~T.~Tanaka\altaffilmark{1}, \L .~Stawarz\altaffilmark{2,\,3}, J.~Finke\altaffilmark{4}, C.~C.~Cheung\altaffilmark{4}, C.~D.~Dermer\altaffilmark{4}, J.~Kataoka\altaffilmark{5},\\ A.~Bamba\altaffilmark{6}, G.~Dubus\altaffilmark{7}, M.~De Naurois\altaffilmark{8}, S.~J.~Wagner\altaffilmark{9}, Y.~Fukazawa\altaffilmark{10}, and D.~J. Thompson\altaffilmark{11}}

\email{ytanaka@hep01.hepl.hiroshima-u.ac.jp}

\altaffiltext{1}{Hiroshima Astrophysical Science Center, Hiroshima University, 1-3-1 Kagamiyama, Higashi-Hiroshima 739-8526, Japan}
\altaffiltext{2}{Institute of Space and Astronautical Science, JAXA, 3-1-1 Yoshinodai, Chuo-ku, Sagamihara, Kanagawa 252-5210, Japan}
\altaffiltext{3}{Astronomical Observatory, Jagiellonian University, ul. Orla 171, 30-244 Krak\'ow, Poland}
\altaffiltext{4}{Space Science Division, Naval Research Laboratory, Washington, DC 20375-5352, USA}
\altaffiltext{5}{Research Institute for Science and Engineering, Waseda University, 3- 4-1, Okubo, Shinjuku, Tokyo 169-8555, Japan}
\altaffiltext{6}{Department of Physics and Mathematics, Aoyama Gakuin University, 5-10-1 Fuchinobe, Chuo, Sagamihara, Kanagawa 252-5258, Japan}
\altaffiltext{7}{UJF-Grenoble 1 / CNRS-INSU, Institut de Plan\'{e}tologie et d'Astrophysique de Grenoble (IPAG) UMR 5274, Grenoble, F-38041, France}
\altaffiltext{8}{Laboratoire Leprince-Ringuet, Ecole Polytechnique, CNRS/IN2P3, F-91128 Palaiseau, France}
\altaffiltext{9}{Landessternwarte, Universit\"at Heidelberg, K\"onigstuhl, D-69117 Heidelberg, Germany}
\altaffiltext{10}{Department of Physical Sciences, Hiroshima University, Higashi-Hiroshima, Hiroshima 739-8526, Japan}
\altaffiltext{11}{NASA Goddard Space Flight Center, Greenbelt, MD 20771, USA}

\begin{abstract}
We report on our study of high-energy properties of two peculiar TeV emitters: the ``extreme blazar" \newhs\ and the ``extreme blazar candidate" \hess\ located near the Galactic Plane. Both objects are characterized by quiescent synchrotron emission with flat spectra extending up to the hard X-ray range, and both were reported to be missing GeV counterparts in the \fermi\ 2--year Source Catalog. We analyze a 4.5 year accumulation of the \fermi\ data, resulting in the detection of \newhs\ in the GeV band, as well as in improved upper limits for \hess. We also present the analysis results of newly acquired \suz\ data for \hess. The X-ray spectrum is well represented by a single power law extending up to 25\,keV with photon index $2.00 \pm 0.02$ and a moderate absorption in excess of the Galactic value, in agreement with previous X-ray observations. No short-term X-ray variability was found over the 80\,ks duration of the \suz\ exposure. Under the blazar hypothesis, we modeled the spectral energy distributions of \newhs\ and \hess, and derived constraints on the intergalactic magnetic field strength and source energetics. We conclude that although the classification of \hess\ has not yet been determined, the blazar hypothesis remains the most plausible option, since in particular the broad-band spectra of the two analyzed sources along with the source model parameters closely resemble each other, and the newly available \wise\ and \uki\ data for \hess\ are consistent with the presence of an elliptical host at the distance of approximately $\sim 600$\,Mpc.
\end{abstract}


\section{Introduction}
\label{sec-intro}

A rich population of very high energy (VHE; $E > 100$\,GeV) $\gamma$-ray emitters has been discovered during a systematic scan of the Galactic plane with the High Energy Stereoscopic System \citep[H.E.S.S.;][]{hess-scan}. The majority of these sources are Galactic in origin and those extended beyond the $\sim 0.1^{\circ}$ point-spread function (PSF) of the instrument include supernova remnants (SNRs), evolved pulsar wind nebulae (PWNe), or molecular clouds. H.E.S.S. sources in the Galactic plane that appear point-like are associated with the Galactic high-mass X-ray binaries (HMXBs), or young PWNe (e.g., the Crab Nebula). Outside of the Galactic plane, the unresolved $\gamma$-ray sources have been predominantly identified with active galactic nuclei (AGN). So far, more than 50 AGN, mostly blazars of the BL Lacertae type (hereafter BL Lacs), have been detected in the VHE range\footnote{{\scriptsize \texttt{http://tevcat.uchicago.edu}}}.

The unresolved, steady, and soft-spectrum (photon index $\Gamma_{\rm VHE} = 3.1 \pm 0.5$) TeV source \hess\ was initially discovered in the Galactic plane scan data collected between 2005 and 2008, at the significance level of $7.9 \sigma$ corresponding to a $>0.47$\,TeV photon flux of $\simeq 10^{-12}$\,ph\,cm$^{-2}$\,s$^{-1}$ \citep{HESS2011}. It is located within the $4.4'$ error circle of an unidentified hard X-ray \integral\ source IGR~J19443+2117, which was also seen with {\it ROSAT}, \chandra , and \swift\ \citep{Tomsick09,Landi09,Cusumano10}. Based on the gathered multiwavelength data, \citet{HESS2011} argued in favor of the extragalactic (and in particular blazar) nature of the source \citep[but see][and the discussion below]{Gabanyi13}.

Blazars are AGN with relativistic jets pointed close to the Earth's line of sight. The typical spectral energy distribution (SED) of a blazar is dominated by the non-thermal Doppler-boosted jet emission, and is characterized by two distinct components, or humps in the $\nu F_{\nu}- \nu$ representation: the low-energy one, commonly interpreted as due to synchrotron emission of ultra-relativistic jet electrons, peaking in the infrared--to--X-ray range, and the high-energy one peaking in the $\gamma$-ray regime, most widely believed to be due to inverse-Compton (IC) scattering of low-energy photons by the synchrotron-emitting electrons. The peak energy of both spectral components was found to anti-correlate with the total radiative power, and also with the `Compton dominance', i.e. the ratio of the IC and synchrotron luminosities \citep{Fossati98,Finke13}, although the origins of these correlations are controversial \citep[e.g.,][]{ghisellini98,ghisellini08,giommi12,giommi13}. High-frequency peaked BL Lacs (HBLs) occupy the lowest luminosity/highest frequency end of the blazar luminosity sequence, with the two spectral components peaking in X-rays (typically $0.1-1$\,keV) and in the VHE range ($0.1-1$\,TeV), respectively.

As discussed in \citet{HESS2011}, the broad-band spectral properties of HESS J1943+213 are consistent with the source being an example of the so-called `extreme HBL' \citep[see][]{Costamante01}, only viewed through the Galactic disk and located at a minimum distance of $\sim 600$\,Mpc \citep{HESS2011}. This distance limit would then set the X-ray luminosity of the source as $\gtrsim 10^{45}$\,erg\,s$^{-1}$, which is relatively high for an HBL. Yet what is the most remarkable for this blazar candidate is its hard X-ray spectrum with no apparent cut-off up to 195\,keV photon energies, as indicated by a long accumulation of the {\it Swift}-BAT data \citep[see][]{Baumgartner13}. This spectrum --- if synchrotron in origin, as expected for an HBL --- would then imply the action of persistent and extremely efficient energy dissipation processes, accelerating jet electrons up to the highest accessible energies. In this respect, \hess\ would in fact outshine the most extreme confirmed HBLs known to date, such as \newhs\ \citep{NewHESS} \hs\ \citep{Tavecchio09,Kaufmann11}. We note that blazars are characterized by substantial broad-band variability at all accessible timescales, but the apparent lack of any flux changes in the case of \hess, even though puzzling, does not strictly exclude the blazar hypothesis.

The class of extreme HBLs, which seems peculiar for its physical properties \citep[particularly low magnetization but high bulk velocities of the emitting regions; see][]{Tav10}, may be relevant in the cosmological context, as it was argued that such objects can be utilized to derive lower limits on the intergalactic magnetic field \citep[IGMF; e.g.,][]{Dubus07,Neronov10,Tavecchio11,Dermer11}. That is because the $>$\,TeV photons emitted from distant extreme HBLs are expected to efficiently pair-create on the extragalactic background light (EBL), leading to the formation of electromagnetic cascades evolving at cosmological scales and reprocessing the primary VHE blazar emission down to the GeV range. Since the resulting spectral shape of the reprocessed $\gamma$-ray continuum depends on the magnetization of the intergalactic medium, a precise characterization of the GeV spectra of extreme HBLs --- which are typically very weak GeV emitters \citep{Tavecchio11} --- are crucial in such attempts to constrain the IGMF. 

In this paper we analyze the newly acquired \suz\ and {\it Fermi} Large Area Telescope (LAT) data for \hess. Together with archival infrared data from the Wide-field Infrared Survey Explorer (\wise) and UKIRT Infrared Deep Sky Survey (\uki) for the putative counterpart, we confront its broad-band spectral properties with those of the well-established extreme HBLs, including \newhs\ for which the \fermi\ and infrared data are similarly examined in detail. The data analysis and the analysis results are given in \S~\ref{sec-data} and \S~\ref{sec-results}, respectively. Discussion regarding the nature of \hess\ in the framework of the blazar scenario, along with the modeling of the broad-band spectra for both analyzed targets (including cascade components for different values of the IGMF), are described in \S~\ref{sec-discussion}. Final conclusions are given in \S~\ref{sec-concl}. In the analysis we assume standard cosmology with $H_0=71$\,km\,s$^{-1}$\,Mpc$^{-1}$, $\Omega_m=0.27$, and $\Omega_{\Lambda}=0.73$.

\section{Data and Data Reduction}
\label{sec-data}

\subsection{\hess}
\label{hs-data}

\subsubsection{\fermi\ Data}
\label{sec-LAT}

The LAT is a pair-production telescope onboard the \textit{Fermi} satellite with large effective area (6500\,cm$^2$ on axis for $>1$\,GeV photons) and large instantaneous field of view (2.4\,sr at 1\,GeV), and is sensitive to $\gamma$ rays from $20$\,MeV to $> 300$\,GeV \citep{LAT}. Here, we analyzed \fermi\ data for both \hess\ and \newhs\ using the Fermi Science Tools version {\tt v9r27p1}. The LAT data were accumulated from 2008 August 4 to 2013 February 08 and we selected 10--300\,GeV \texttt{SOURCE} class events using \texttt{gtselect}. Regions of Interest (RoIs) were set to $10^{\circ}$ circular regions centered at each source position. To eliminate Earth limb $\gamma$ rays, the maximum zenith angle was set to 100$^{\circ}$. Good-quality and science-configurated LAT data were selected from standard all-sky data by using \texttt{gtmktime}. A 50$^{\circ}$ rocking angle cut is also applied. We performed unbinned likelihood analysis using \texttt{gtlike} and utilized the \texttt{P7SOURCE\_V6} instrument response functions. In the XML source model, we included 2FGL sources \citep{Nolan12} within a $10^{\circ}$ circular region assuming their power-law spectra with free photon indices and normalizations. Spectral parameters of 2FGL sources within an annulus of 10--15$^{\circ}$ were fixed to the 2FGL values and included in the source model, together with \texttt{gal\_2yearp7v6\_v0.fits} and \texttt{iso\_p7v6source.txt} which represent the Galactic and isotropic diffuse background emissions, respectively. Here, the photon index of the power-law scaling of the Galactic diffuse template is fixed to 0. Data points of the GeV spectra were calculated by repeating \texttt{gtlike} under assumption of a single power-law spectral shape in each energy range.\footnote{{\tiny \texttt{likeSED.py}, available at \texttt{http://fermi.gsfc.nasa.gov/ssc/data/analysis/user/}}}

\subsubsection{Archival X-ray Data}
\label{sec-archiv}

\hess\ is located within the 4.4\,arcmin error circle of the unidentified hard X-ray \integral-IBIS source IGR~J19443+2117, with a centroid about 3.2\,arcmin offset from the H.E.S.S. position. The soft X-ray counterpart to IGR~J19443+2117 was detected by \chandra\ \citep[CXOU J194356.2+211823;][]{Tomsick09} and \swift\ \citep[SWIFT J1943.5+2120;][]{Landi09}, with exposure times of 4.8\,ks and 11\,ks, respectively. The same X-ray source was also detected in the past by {\it ROSAT}-HRI (1RXH J194356.2+211824). The combined power-law fit to \swift-XRT and \integral-IBIS data returned the photon index of $\Gamma_{\rm X} = 2.04\pm0.12$ with the energy fluxes $F_{\rm 2-10 keV}= \left( 1.83\pm0.04 \right) \times 10^{-11}$\,erg\,cm$^{-2}$\,s$^{-1}$ and $F_{\rm 20-100 keV} = \left( 1.12\pm0.22 \right) \times 10^{-11}$\,erg\,cm$^{-2}$\,s$^{-1}$ \citep{HESS2011}. Within the error of the cross-calibration constant between XRT and IBIS, $\simeq 0.60$, the IBIS data are in agreement with the XRT spectrum. The \chandra\ observations (performed two years earlier, in 2008) are consistent with the \swift\ results as well, with the derived energy flux $F_{\rm 0.3-10 keV} = \left( 2.9^{+2.4}_{-0.5} \right) \times 10^{-11}$\,erg\,cm$^{-2}$\,s$^{-1}$ and $\Gamma_{\rm X}=1.83\pm0.11$. See Fig.~7 of \citet{HESS2011} for the consistency of the X-ray fluxes between \swift\ and \chandra. The \integral-SPI observations provided the upper limit for the source flux $F_{\rm >100 keV} < 2.6 \times 10^{-11}$\,erg\,cm$^{-2}$\,s$^{-1}$ \citep{Bouchet08}. Finally, the hard X-ray counterpart of \hess/ IGR~J19443+2117 is also present in the 54-month Palermo \swift-BAT Catalog \citep[PBC J1943.9+2118, $F_{\rm 14 -150 keV} = \left( 2.2\pm0.7 \right) \times 10^{-11}$\,erg\,cm$^{-2}$\,s$^{-1}$;][]{Cusumano10}, and in the 70 month \swift-BAT catalog \citep{Baumgartner13}. The BAT light curve seems to reveal some hints for flux changes over 70 months of monitoring\footnote{{\tiny \texttt{http://swift.gsfc.nasa.gov/results/bs70mon/SWIFT\_J1943.5p2120}}}, but low photon statistics precludes a detailed analysis of the source variability.

A power-law model fit to the spectrum provided by \citet{Baumgartner13} yielded $\Gamma_{\rm X} = 2.12^{+0.25}_{-0.24}$ and the energy flux $F_{\rm 14 -195 keV} = 2.83^{+0.47}_{-0.44} \times 10^{-11}$\,erg\,cm$^{-2}$\,s$^{-1}$. There is no evidence for any high-energy cut-off in the BAT spectrum. Also, even though the IBIS source is weak, it appears steady with no signs of any large-amplitude variability. The hydrogen column densities derived by means of modeling the \chandra\ and \swift\ spectra were $N_{\rm H} \simeq (1.89^{+0.25}_{-0.22}) \times 10^{22}$\,cm$^{-2}$ \citep{Tomsick09} and $(1.37^{+0.12}_{-0.13}) \times 10^{22}$\,cm$^{-2}$ \citep{Landi09}, respectively.\footnote{{\scriptsize It is unclear why the \swift-XRT 11 ks observation provided much smaller uncertainty when compared to the \chandra\ 4.8 ks observation, taking into account the fact that \chandra\ ACIS-I has about four times larger effective area with respect to \swift-XRT at 1 keV \citep[see Figure~1 in][]{Tundo12}.}} In spite of this column density being in excess of the expected Galactic one in the direction of the source based on the \ion{H}{1} mapping (see \S~\ref{sec-X} below), no conclusive evidence for the intrinsic absorption in the X-ray spectrum of \hess\ was presented in the past.

\subsubsection{New \suz\ Observation}
\label{sec-suz}

We observed \hess\ with the \suz\ X-ray satellite \citep{Mitsuda2007} on 2011 November 10--11 (Obs ID 706007010). We analyzed the data taken with both X-ray CCD cameras onboard, namely the X-ray Imaging Spectrometer \citep[XIS;][]{Koyama2007} and Hard X-ray Detector \citep[HXD;][]{Takahashi2007,Kokubun2007}. Currently, three CCDs are working well, two of which are front-illuminated (XIS0 and XIS3) and the other one is back-illuminated (XIS1). The HXD consists of a Si semiconductor detector (PIN) and the GSO scintillator, sensitive to hard X-rays in the ranges of $\sim 10-50$\,keV and $\sim 50-600$\,keV, respectively. Here we did not analyze the GSO data because the relatively small flux of the targeted source does not allow us to detect hard X-rays above $\sim 50$\,keV over the high background.

We used the XIS data processing script version 2.7.16.31 and followed the screening criteria described in {\it The Suzaku Data Reduction Guide}\footnote{{\scriptsize \texttt{http://heasarc.nasa.gov/docs/suzaku/analysis/abc/}}}. Events with \texttt{GRADE} 0, 2, 3, 4 and 6 were utilized in the data reduction procedure, and flickering pixels were removed by using \texttt{cleansis}. We selected good-time intervals by applying \texttt{SAA\_HXD}$==$0 $\&\&$ \texttt{T\_SAA\_HXD}$>$436 $\&\&$ \texttt{ELV}$>$5 $\&\&$ \texttt{DYE\_ELV}$>$20 $\&\&$ \texttt{ANG\_DIST}$<$1.5 $\&\&$ \texttt{S0\_DTRATE}$<$3 $\&\&$ \texttt{AOCU\_HK\_CNT3\_NML\_P}$==$1. After the event screening, the net exposure was 38.9 ks. In the analysis we used the \texttt{HEADAS} software version 6.12 and calibration database (CALDB) version released on 2012 Feb 11. 

\subsubsection{Lower Frequencies}
\label{sec-low}

There is only one radio source in the NVSS catalog within the error circle of HESS J1943+213, namely NVSS J194356+211826, offset by 14.7\,arcsec from the H.E.S.S. source position and 3.5\,arcsec from the \chandra\ source position. This radio source (1.4\,GHz flux density of $0.103$\,Jy) has been detected in various survey programs and dedicated exposures between 327 and 4850\,MHz; no evidence for flux variations over the 12-year time span of the collected radio data was found \citep{HESS2011}. Recent high-resolution data taken with the European VLBI Network (EVN) also revealed a compact radio source, and the low brightness temperature is claimed to be an argument against the blazar origin for this object \citep{Gabanyi13}.

A faint unidentified near-infrared counterpart of \hess\ was also found in 2MASS data in the $K$ band, 2MASS J19435624+2118233, well within the small \chandra\ error circle; at smaller wavelengths (2MASS $J$ and $H$ bands, as well as \swift-UVOT $V$ band), only upper limits for the source flux were found \citep{HESS2011}. We examined the subsequently available all-sky mid-infrared (MIR) data provided by \wise\ satellite \citep{Wright10} in four MIR bands, W1, W2, W3, and W4, centered on wavelengths around 3.4, 4.6, 12, and 22\,$\mu$m, respectively. The PSF of the telescope corresponds to a Gaussian of about 6.1, 6.8, 7.4, and 12\,arcsec in W1--W4 respectively, sampled at 2.8, 2.8, 2.8 and 5.6\,arcsec/pix. With nominal 5$\sigma$ point source sensitivities of $\sim$0.08, 0.1, 1 and 6\,mJy in the four bands, this survey is orders of magnitude more sensitive than previous infrared all sky surveys\footnote{{\scriptsize \texttt{http://wise2.ipac.caltech.edu/docs/release/allsky}}}. The MIR emitter, WISE J194356.25+211823.2, coinciding with \hess, is detected at $3.4$\,$\mu$m, $4.6$\,$\mu$m, and $12$\,$\mu$m wavelengths, with the observed fluxes of $1.88\pm0.19$\,mJy, $1.67\pm0.14$\,mJy, and $1.36\pm0.27$\,mJy, respectively, and is not detected at $22$\,$\mu$m. Using the \citet{Schlegel98} coefficients in adjacent bands, we can also estimate extinction corrected flux densities of $3.1$\,mJy ($3.4$\,$\mu$m) and $1.9$\,mJy ($4.6$\,$\mu$m); extinction is negligible at $12$\,$\mu$m. The 2MASS source is also catalogued in the \uki-DR6 Galactic plane survey \citep{Lucas08} as UGPS J194356.23+211823.3, with observed $J=16.448\pm0.010$, $H=15.187\pm0.006$, and $K=14.174\pm0.006$ mag. Using the \citet{Schlegel98} extinction corrections adopted in \citet{HESS2011}, these correspond to the extinction-corrected fluxes of $4.18\pm 0.04$\,mJy ($1.25$\,$\mu$m), $3.71\pm 0.02$\,mJy ($1.65$\,$\mu$m), and $3.60\pm 0.02$\,mJy ($2.21$\,$\mu$m), respectively. New near-infrared data for \hess\ will be presented in the forthcoming paper by \citep{Peter14}.

\subsection{\newhs}
\label{newhs-data}

The \fermi\ data selection and analysis for \newhs\ are the same as in the case of \hess\ (see \S~\ref{sec-LAT}), except for the 1--300 GeV energy range considered. The reason why we decreased the lower energy threshold to 1 GeV is that \newhs\ is clearly detected in 10--300 GeV \texttt{SOURCE} class data. At lower frequencies, in addition to the archival \swift-UVOT and XRT, and ATOM (host galaxy-subtracted) data discussed in \citet{NewHESS}, we also include the newly updated \swift-BAT spectrum of the source \citep{Baumgartner13}, and the most recent infrared measurements summarized as follows: \newhs\ is detected in the \wise\ data at $3.4$\,$\mu$m, $4.6$\,$\mu$m, and $12$\,$\mu$m wavelengths, yielding the observed fluxes of $0.64\pm0.02$\,mJy, $0.57\pm0.02$\,mJy, and $0.40\pm0.11$\,mJy, respectively, with negligible extinction in these bands; the extinction-corrected 2MASS fluxes for the source are $0.69\pm0.06$\,mJy ($1.25$\,$\mu$m), $0.61\pm0.07$\,mJy ($1.65$\,$\mu$m), and $0.93\pm0.09$\,mJy ($2.17$\,$\mu$m).

\section{Results}
\label{sec-results}

\subsection{X-ray Spectrum of \hess}
\label{sec-X}

\begin{figure}
\begin{center}
\plotone{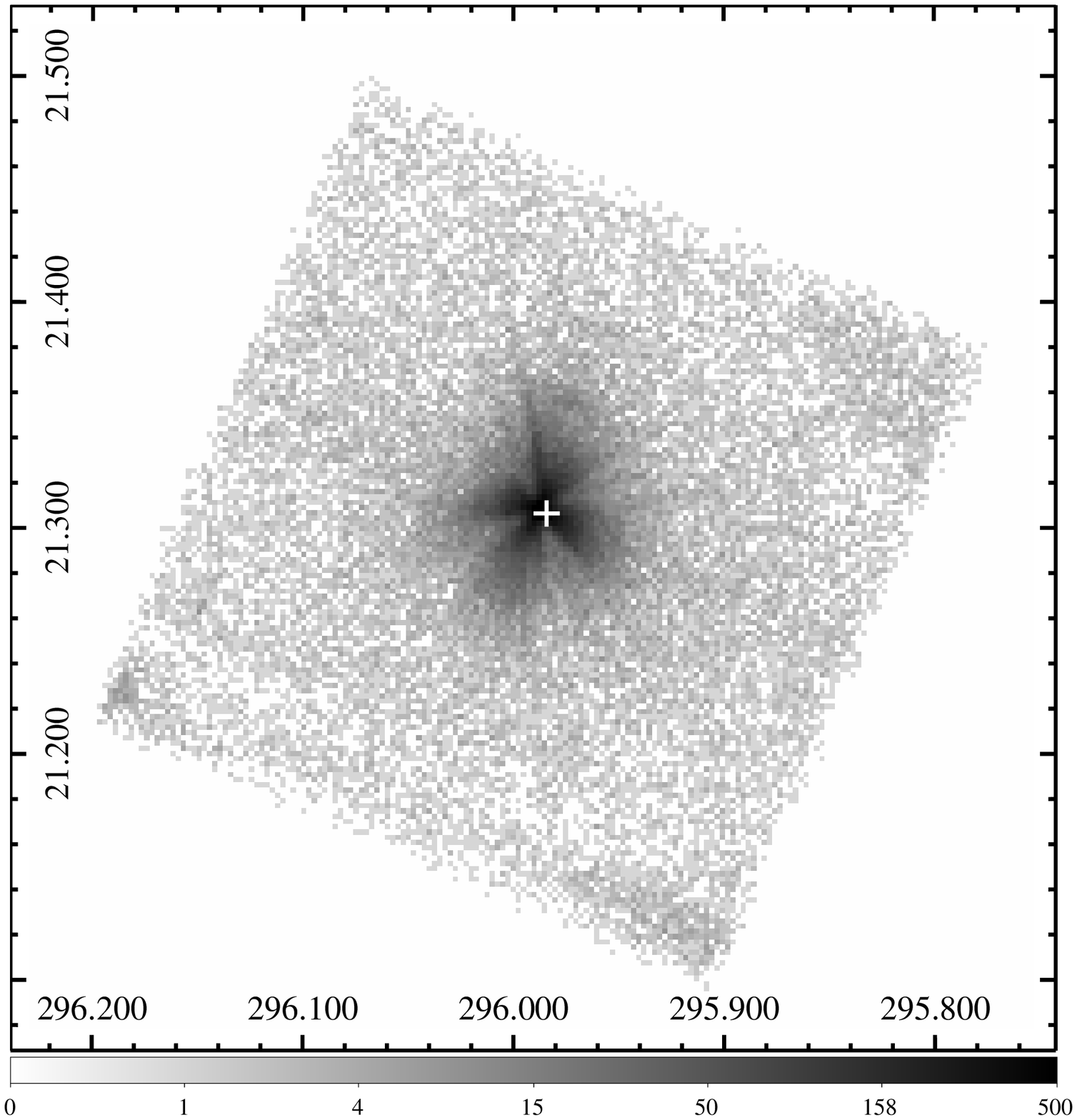}
\caption{\suz\ XIS0+3 image of \hess\ with a pixel scale of 8.32$^{\prime\prime}$ pixel$^{-1}$. Horizontal and vertical axes are R.A. (J2000) and Dec. (J2000), respectively, in units of degrees. White cross denotes the source position measured by \chandra. The positions of the NVSS, EVN, \wise, and 2MASS counterparts overlap with the \chandra\ source localization, and hence are not shown here. A slight offset between the \suz\ and \chandra\ positions is within the $\lesssim 1$\,arcmin \suz\ pointing accuracy.}
\label{fig-suzim}
\end{center}
\end{figure}

\begin{figure}
\begin{center}
\plotone{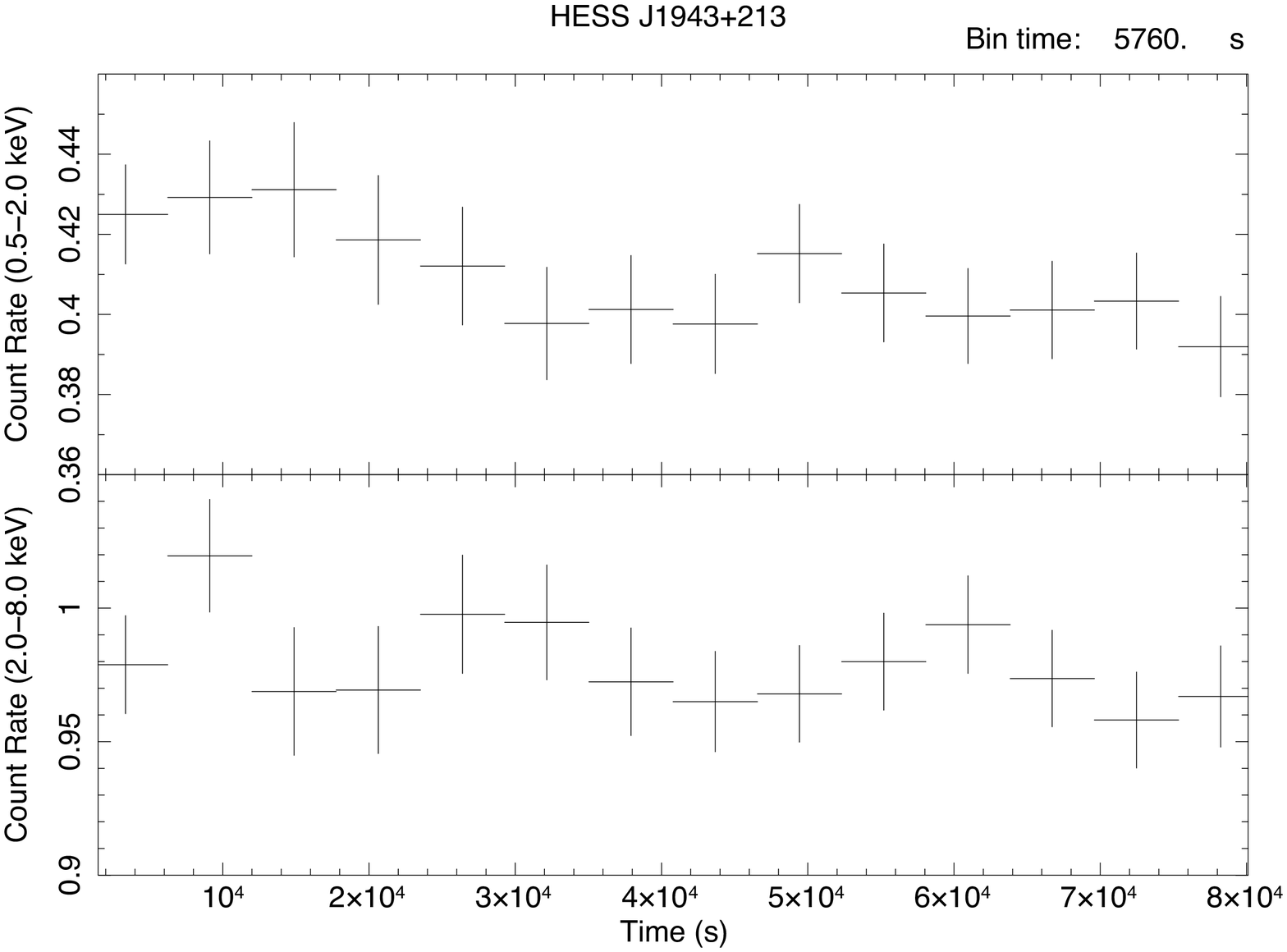}
\caption{Background-subtracted \suz\ XIS0+3 light curve of \hess\ with 5760 s bins (Upper panel: $0.5-2.0$\,keV, Lower panel: $2.0-8.0$\,keV).}
\label{fig-suzlc}
\end{center}
\end{figure}

The \suz-XIS0+XIS3 combined image and light curves (in the $0.5-2.0$ and $2.0-8.0$\,keV bands) of \hess\ are shown in Figures~\ref{fig-suzim} and \ref{fig-suzlc}, respectively. No significant variability or modulation was seen over the entire 80\,ks \suz\ pointing (40\,ks net exposure). Since the hard X-ray signal detected with HXD/PIN was not very strong, the resulting light curve of the source above 10\,keV was partly affected by statistical and systematic fluctuations, and as such is rather inconclusive. We also did not find any source extension beyond the XIS PSF of $\simeq 1.5$\,arcmin.

\begin{figure}
\begin{center}
\includegraphics[scale=0.3]{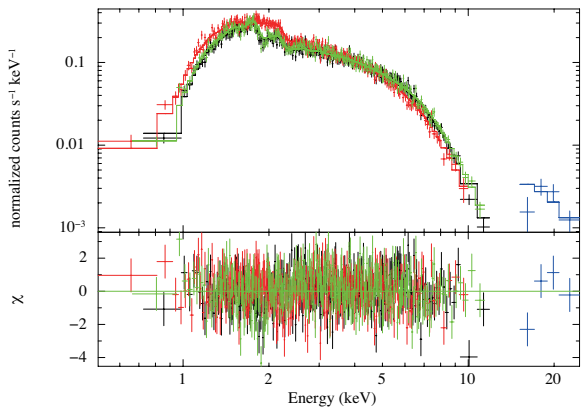}
\caption{\suz\ spectrum of \hess\ in the $0.5-25$\,keV range obtained with XIS and HXD/PIN. Black, red, and green crosses denote the XIS0, XIS1, and XIS3 data, respectively, together with PIN data shown in blue. The best power-law model fit, discussed in \S~\ref{sec-X}, is also displayed as black, red, green and blue curves.}
\label{fig-suzspec}
\end{center}
\end{figure}

Figure~\ref{fig-suzspec} presents the simultaneous X-ray spectrum of the target within the $0.5-25$\,keV range (combined \suz-XIS and HXD/PIN data). The spectrum is well represented by a single power-law model (reduced $\chi^2_{\nu}$/d.o.f $=$ 1.11/847) with photon index $\Gamma=2.00\pm0.02$, absorbed by the hydrogen column density of $N_{\rm H}= \left(1.38\pm0.03 \right) \times 10^{22}$\,cm$^{-2}$, in agreement with the one emerging from the previous \swift\ observations, but in excess over the Galactic value of $8.37 \times 10^{21}$ cm$^{-2}$ estimated based on the Galactic \ion{H}{1} Leiden/Argentine/Bonn (LAB) Survey\footnote{{\tiny \texttt{http://heasarc.gsfc.nasa.gov/cgi-bin/Tools/w3nh/w3nh.pl}}} \citep{Kalberla2005}. We also applied an unabsorbed broken power-law model to the \suz\ spectrum of \hess\, and rejected it because of a much larger reduced chi-square value of 1.21 with respect to the single absorbed power-law model. The unabsorbed $2-10$\,keV flux of the source reads as $\left(1.85^{+0.07}_{-0.06} \right) \times 10^{-11}$\,erg\,cm$^{-2}$\,s$^{-1}$, again in remarkable agreement with the previous \swift-XRT observations \citep{HESS2011}, indicating that \hess\ is a truly steady X-ray emitter.

In order to investigate in more detail the issue of the excess absorption in the X-ray spectrum of \hess, we have modeled the acquired \suz\ data with the \texttt{SYNCHROTRON} code developed and implemented in {\tt XSPEC} by \citet{Ushio09,Ushio2010}. In this ``parametric forward-fitting'' model, the observed X-ray emission of a source is fitted with the synchrotron continuum originating from a given (assumed) shape of the electron energy distribution, instead of the assumed spectral form of the non-thermal emission continuum. Therefore, the {\tt XSPEC} fitting with the \texttt{SYNCHROTRON} model allows us to check in particular if the suppressed soft X-ray flux of \hess\ is due to the excess absorption indeed, or is rather due to the high low-energy cut-off in the underlying electron energy distribution \citep[see in this context the case of \hs;][]{Tavecchio09,Kaufmann11}. In order to reduce the number of model free parameters, in the fitting procedure we assumed a single power-law electron energy distribution $dN/d\gamma \propto \gamma^{-s}$ with fixed energy index $s=3.0$ and the maximum electron Lorentz factor $\gamma_{max} = 10^8 /\sqrt{\delta \, B' / 0.1\,{\rm G}}$ (where $\delta$ is the Doppler factor of the emission region, and $B'$ is the comoving magnetic field intensity); only the minimum electron Lorentz factor $\gamma_{min}$ and the absorbing column density $N_{\rm H}$ were allowed to vary. The modeling returned  $\gamma_{min} = 10^{5.8^{+0.02}_{-0.05}} /\sqrt{\delta \, B' / 0.1\,{\rm G}}$, and $N_{\rm H}= \left(1.23^{+0.06}_{-0.04} \right) \times 10^{22}$\,cm$^{-2}$ still in excess of the Galactic value, even though the observed flux decrease below 1\,keV was in this case ascribed partly to the high low-energy electron cut-off. We note that the analogous modeling with $N_{\rm H}$ fixed at the Galactic value resulted in a large reduced chi-square, while on the other hand the modeling with frozen $\gamma_{min} = 10^5 /\sqrt{\delta \, B' / 0.1\,{\rm G}}$ returned $s=3.03^{+0.04}_{-0.05}$ and $N_{\rm H}= \left(1.39 \pm 0.03 \right) \times 10^{22}$\,cm$^{-2}$ in agreement with the emission continuum power-law fit discussed in the previous paragraph. Hence, we conclude that the signatures for the excess absorption in the X-ray spectrum of \hess, even though on a rather modest level, can nonetheless be claimed robustly.

We note that previously some evidence for the intrinsic absorption of the X-ray spectra of HBLs has been presented only in a few cases, namely Mrk\,501 \citep{Kataoka99,LAT501}, and more recently 1RXS J101015.9--311909 \citep{1010}.

\subsection{GeV Emission of \hess}
\label{sec-GeV}

\begin{figure}
\begin{center}
\plottwo{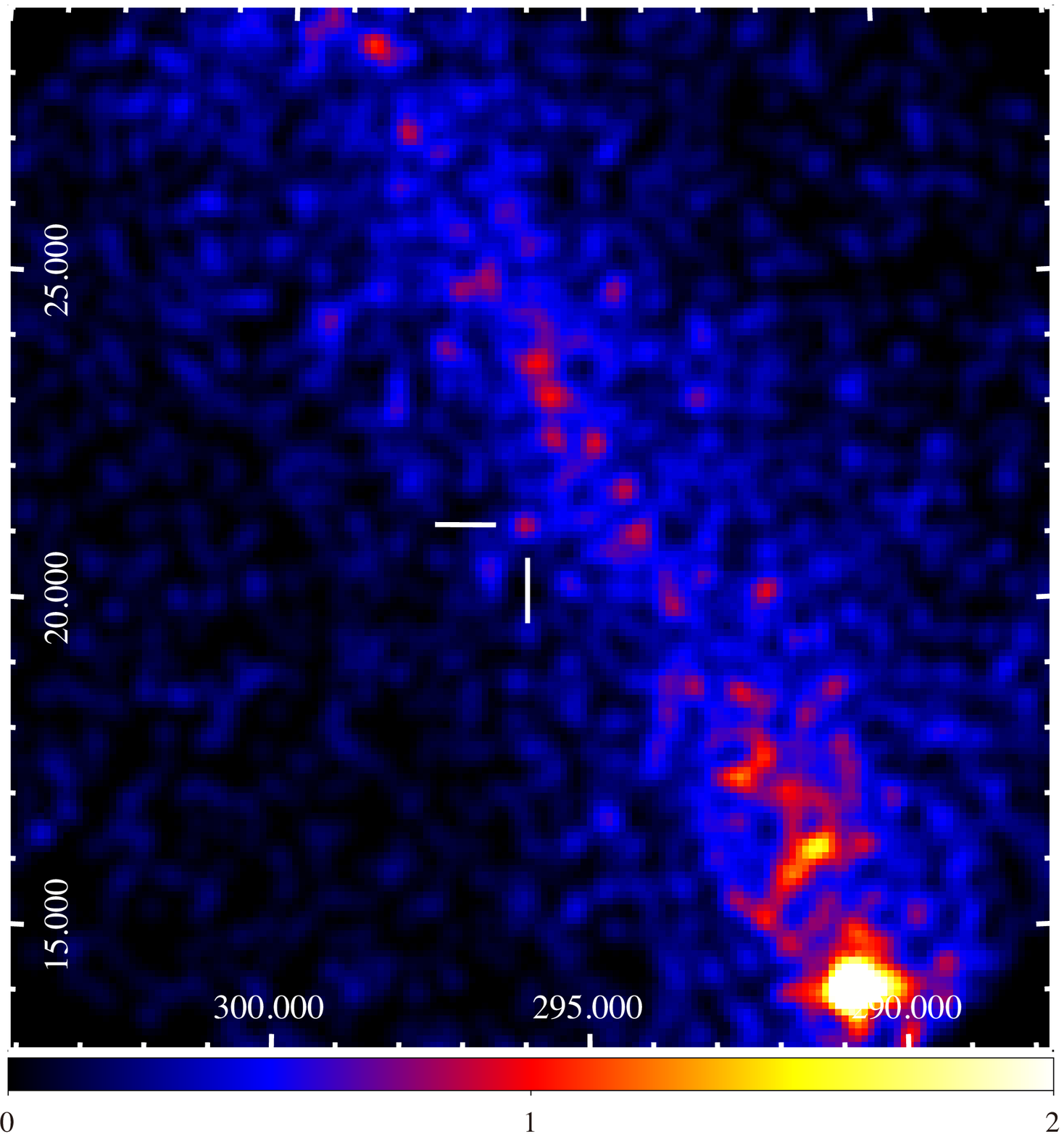}{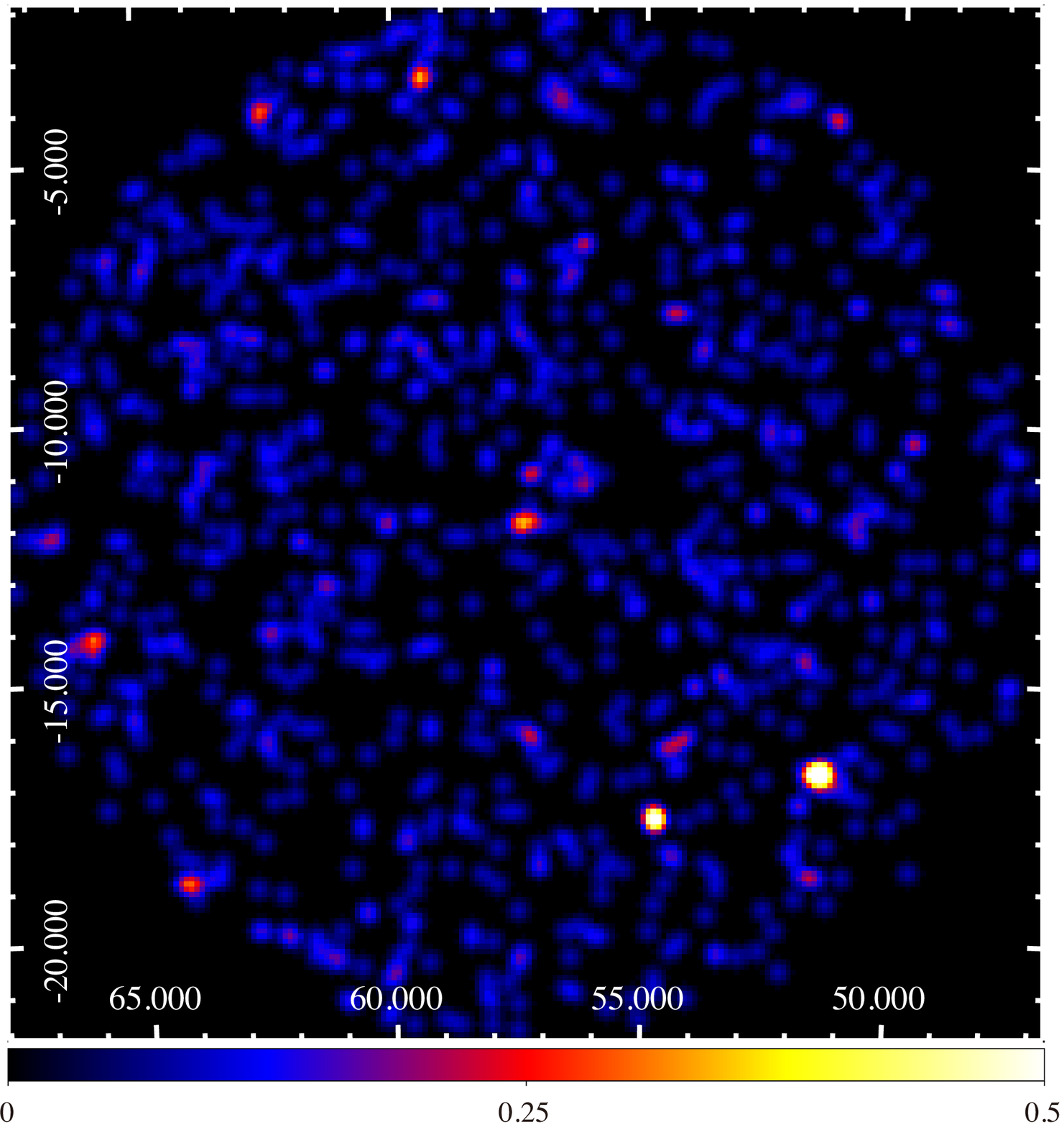}
\caption{{\it Left}: \fermi\ 10--300 GeV count map centered on the position of \hess\ (R.A.=295.9792$^{\circ}$, Dec.=21.30222$^{\circ}$), which is indicated by white ticks. {\it Right}: Same as the {\it Left} figure but for the 10$^{\circ}$ radius circular region centered on the position of \newhs\ (R.A.=57.3466$^{\circ}$, Dec.=$-$11.9908$^{\circ}$). Both images are smoothed with Gaussian kernel of three pixels (scale: 0.1$^{\circ}$ per pixel). Note that the size of the plotted area is different between the two images ({\it Left} figure is more zoomed).}
\label{fig-lat1943}
\end{center}
\end{figure}

Figure~\ref{fig-lat1943} shows the \fermi\ count map of the $10^{\circ}$ radius circular region centered on \hess\ in the $10-300$\,GeV photon energy range. A weak GeV counterpart seems to be present at the position of \hess\ ($l =57.764^{\circ}$, $b=-1.295^{\circ}$ in Galactic coordinates), but similar faint point-like enhancements are also seen along the Galactic plane, where the intense foreground Galactic diffuse emission dramatically dilutes any signal from $\gamma$-ray emitters located in the plane. 
In order to minimize the number of degrees of freedom, we tested only the hypothesis that there is a LAT source at the HESS position.
We investigated the detection significance using \texttt{gtlike} following the analysis flow described in \S2.3 and obtained the Test Statistic (TS) value of 22.3, which is just below the formal detection threshold of TS\,=\,25 (corresponding to a $\sim 5 \sigma$ significance). Note that the TS is defined as $-2 \left( \ln L_0 - \ln L_1 \right)$, where $L_0$ and $L_1$ are the likelihood values with and without the source, respectively \citep[see][]{Mattox96}. Here we cannot exclude a possibility for the detection of a spurious source coincident with \hess\ simply due to subtraction of the imperfect template for the Galactic diffuse emission. In fact, the presence of several similar faint point-like features along the Galactic plane in Figure~\ref{fig-lat1943} may suggest that this is indeed the case. In addition, assuming the source detection, the derived power-law index of the 10--300\,GeV counterpart of \hess\ is $\Gamma \sim 2.4$, which is suspiciously soft, although this may indicate a spectral turnover around that energy range. Analysis of newly released Pass 7 reprocessed LAT data\footnote{{\tiny available at \texttt{http://fermi.gsfc.nasa.gov/cgi-bin/ssc/LAT/LATDataQuery.cgi}}} accumulated over 5 years shows a detection of \hess\ with a rather flat spectrum ($\Gamma \sim 1.6$) at energies above 1 GeV \citep{Peter14}, consistent with our suggestion of a spectral turnover at higher energies. Given all these cautions and limitations, instead of claiming the detection here we provide only the 95\% flux upper limits for \hess: $F_{\rm 10-30\,GeV} < 2.5 \times 10^{-12}$\,erg\,cm$^{-2}$\,s$^{-1}$, $F_{\rm 30-100\,GeV} < 4.1 \times 10^{-12}$\,erg\,cm$^{-2}$\,s$^{-1}$, and $F_{\rm 100-300\,GeV} < 2.7 \times 10^{-12}$\,erg\,cm$^{-2}$\,s$^{-1}$, assuming single power-law spectra with photon indices $\Gamma =2.0$ in each band and using at the point that the $-\log \left( {\rm likelihood} \right)$ increased by 1.36. These are shown in the SED representation in Figure~\ref{fig:SED2}, along with the H.E.S.S. spectrum from \citet{HESS2011}, the infrared (\wise\ and \uki) data discussed in \S~\ref{sec-low}, and finally all the available (archival and the new \suz) X-ray spectra of the source introduced and analyzed in \S~\ref{sec-archiv} and \ref{sec-X}.

\subsection{GeV Emission of \newhs}
\label{sec-newGeV}

In contrast, the only known extreme HBL that was missing thus far its GeV counterpart, \newhs, is now clearly detected in the most recent accumulation of the \fermi\ data. We derived TS\,=\,51.0 and power-law index $\Gamma_{\rm LAT}=1.65 \pm 0.17$ with the 1--300 GeV flux of $\left( 3.0\pm0.7 \right) \times 10^{-10}$ ph cm$^{-2}$ s$^{-1}$ for the source. The rather low GeV flux of the source precluded any variability studies. The broad-band SED of the blazar is given in Figure~\ref{fig:SED2}, including the newly derived LAT data points, the H.E.S.S. spectrum from \citet{NewHESS}, and optical to X-ray data introduced and discussed in \S~\ref{newhs-data}.

\section{Discussion}
\label{sec-discussion}

\begin{figure}
\begin{center}
\plotone{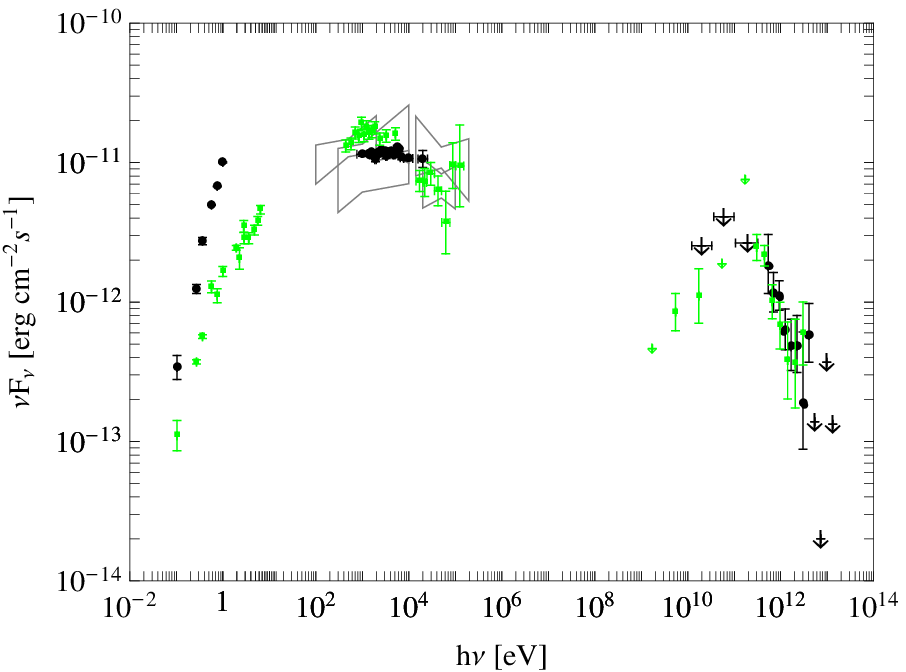}
\caption{Broad-band SEDs of \hess\ and \newhs. Black circles and large arrows correspond to the \hess\ VHE fluxes/upper limits taken from \citet{HESS2011}, as well as the \wise, \uki, \suz\ (re-binned XIS0+PIN), and \fermi\ data derived in this paper; gray bowties denote the archival X-ray spectra of the source (see \S~\ref{hs-data} and \ref{sec-GeV}). Green squares and small arrows correspond to the \newhs\ VHE, \swift-UVOT \& XRT, and ATOM (host galaxy-substracted) data taken from \citet{NewHESS}, the updated \swift-BAT spectrum from \citet{Baumgartner13}, along with the \fermi, \wise, and the 2MASS fluxes derived in this paper (see \S~\ref{newhs-data} and \ref{sec-newGeV}). \label{fig:SED2}}
\end{center}
\end{figure}

The broad-band SEDs of \hess\ and \newhs\ shown in Figure~\ref{fig:SED2} resemble each other closely, although some differences can be noted as well. In particular, the X-ray continuum of \newhs, which is known to vary \citep[see][]{Tavecchio11}, is in general steeper than that of \hess. Also, the IR segments of the spectra are distinct, although this difference seems to be consistent with the idea of a more prominent contribution from the elliptical host in \hess\ due to the smaller distance of the source ($\sim 600$\,Mpc versus $\simeq 900$\,Mpc for \newhs; see the discussion below). The overall correspondence between the two spectra strongly supports the blazar scenario for \hess.

\subsection{Possible identification of \hess}
\label{sec-nature}

The nature of \hess\ is puzzling and the subject of ongoing debate. Three main scenarios identifying the source as a HMXB, a young PWN, or an extreme HBL have been outlined in \citet{HESS2011} and discussed later in \citet{Leahy12} and \citet{Gabanyi13}. Here we briefly revisit this issue in light of the new data.

The flat X-ray spectrum of \hess\ with photon index $\Gamma_{\rm X} = 2.0$ and no high-energy cut-off up to the 100\,keV range, together with the very soft VHE spectrum $\Gamma_{\rm VHE} = 3.2 \pm 0.5$, are consistent with the HMXB hypothesis. The lack of any orbital modulation in the X-ray band might be baffling in this context, but still does not exclude the possibility of the discussed object being a binary system. Similarly, no (or a very weak) GeV counterpart does not contradict the idea \citep[keeping in mind the case of HESS J0632+057; see][]{Caliandro13}. The most problematic aspect of the scenario is, however, the absence of a massive companion manifesting as a pronounce IR/optical source. As discussed in \citet{HESS2011}, the 2MASS $K$-band counterpart of \hess\ could in principle be attributed to the massive O or Be-type star taking into account the 2MASS and  \swift-UVOT upper limits derived at shorter wavelengths, but only in the case of a large distance to the system, $d \simeq 25$\,kpc. This would in turn imply exceptionally high (for an HMXB) X-ray luminosity, exceeding $10^{36}$\,erg\,s$^{-1}$. In addition, the arcmin-scale elongated radio halo of \hess\ discovered by \citet{Gabanyi13} would be similarly in conflict with the binary scenario. The analysis of the \suz\ data presented in this paper does not add any new conclusive piece of evidence to the above discussion, as no orbital modulation has been detected in our new X-ray exposure. However, the \wise\ and especially \uki\ detections of the infrared counterpart of \hess\ discussed in this paper, with the resulting fluxes below the previous upper limits, would push the distance of the presumed O-type stellar companion of the high-energy source even outside the 25\,kpc radius, making the HMXB hypothesis even less plausible.

In the framework of the PWN model, no extension of the source in the H.E.S.S. data together with the VHE/X-ray luminosity ratio $L_{\rm 1-30\,TeV}/L_{\rm 2-10\,keV} \simeq 0.04$ would imply a young age ($\simeq 10^3$\,yr) and a relatively high, Crab-like spin-down power ($\dot{E} \simeq 10^{38}$\,erg\,s$^{-1}$) of the system located within $16$\,kpc \citep{HESS2011}. On the other hand, the soft VHE spectrum and the point-like appearance of the source in the \chandra\ data question to some extent the PWN nature of \hess. In addition, based on the \ion{H}{1} absorption spectrum, \citet{Leahy12} argued for the source distance exceeding 16\,kpc. These objections have been addressed by \citet{Gabanyi13} who, based on the newly acquired EVN and archival VLA data, argued that the discussed object can still be a PWN left over after a supernova explosion that happened in a low-density environment at the distance of $d \simeq 17$\,kpc. The fact that the X-ray counterpart of \hess\ appears point-like for \suz-XIS and unresolved even by \chandra\ is not any strong argument against the PWN hypothesis. However, the presence of a bright flat-spectrum mid- to near-infrared counterpart, as discussed in this paper, is unexpected for a PWN. 

A comparison of high-energy spectral properties between \hess\ and bona fide extreme HBLs is summarized in Table~\ref{table_eHBLs} and discussed in more detail in the next section by means of a spectral model comparison with \newhs. This comparison indicates that the multi-wavelength spectral properties of \hess\ are in principle consistent with it being a member of this class, even taking into account the hard, broad-band X-ray continuum and a very weak GeV emission of the source. The amorphous arcmin-scale radio halo surrounding the target \citep{Gabanyi13} would fit this interpretation as well, since the previous VLA studies revealed the presence of analogous features in several BL Lacs \citep[e.g.,][]{Ulvestad84,Ulvestad86}. The lack of X-ray variability implied by different X-ray pointings spread over several years, including our new \suz\ observations as well as the continuous \swift-BAT monitoring, may however cast doubt on the blazar identification. In addition, \citet{Gabanyi13} noted that the brightness temperature of the \hess\ radio core, $T_{\rm b} \sim 8 \times 10^{7}$ K, is much lower than those typically measured in radio cores of blazars, and BL Lacs in particular. While the steady X-ray emission of the target could be explained by a particular duty cycle of HBLs, which seem to be able to undergo extended periods of quiescence \citep[see in this context, e.g.,][for \hs\ and PKS 0548-322, respectively]{Kaufmann11,Perri07}, the relatively low radio brightness temperature --- if intrinsic to the source instead of being due to a small-angular size scattering cloud on the line-of-sight within the Galaxy \citep[see the discussion in][]{Gabanyi13} --- may indeed be considered as problematic for the blazar hypothesis.  We however note in this context that, on the other hand, no superluminal velocities have been found in the TeV-bright HBLs on milli-arcsec scales, despite several dedicated observational programs \citep[e.g.,][and references therein]{Piner10}, and yet such apparent superluminal velocities are considered to be one of the hallmark properties of blazars in general. Note also that in a recent systematic $\sim$1 milli-arcsecond resolution survey of radio-faint BL Lacs \citep{Liuzzo13} using the Very Long Baseline Array ({\it VLBA}), most are characterized by faint $\sim$10 mJy radio cores with inferred brightness temperatures \citep[see][section 5.4 therein]{kov05} as low as a few $10^8$\,K.

Finally, the newly available \wise\ and \uki\ data for \hess\ are consistent with the spectrum of the expected host galaxy located at a larger distance: this is demonstrated in Figure~\ref{fig:1943SED}, where we plotted the template of a luminous elliptical placed at $600$\,Mpc \citep[bolometric luminosity $L_{\rm 0.1-10\,\mu m} \simeq 7 \times 10^{44}$\,erg\,s$^{-1}$; template taken from][]{Silva98}. As argued below, such a distance would additionally be in accord with the overall energetics of the source in the blazar scenario.

\begin{table*}
\noindent
{\caption[]{High-energy spectral properties of extreme HBLs.
\label{table_eHBLs}}}
\begin{center}
\begin{tabular}{llllc}
\hline \hline
Name & $z$ & $\Gamma_{\rm VHE} \pm {\rm stat} \pm {\rm sys}$ & $\Gamma_{\rm LAT}$ & Quiescent X-ray spectrum$^{\dagger}$ \\
\hline
HESS J1943+213  & ? & $3.10 \pm 0.12 \pm0.12$ & --- & $\Gamma=2.0$ with excess absorption \\
1ES 0229+200       & 0.140 & $2.50 \pm 0.19 \pm 0.10$ & $1.36 \pm 0.25$ & $\Gamma=1.84\pm0.02$ with excess absorption \\
1ES 0347$-$121        & 0.185 & $3.10 \pm 0.23 \pm 0.10$ & $1.65 \pm 0.17$ & $\Gamma=1.82\pm0.03$ with excess absorption \\
1ES 1101$-$232        & 0.186  & $2.94 \pm 0.20$ &  $1.80 \pm 0.31$ & $\Gamma_{low}=2.04\pm0.02$, $\Gamma_{high}=2.32\pm0.02$, $E_{brk}=1.37\pm0.08$\,keV \\
1ES 1218+304       & 0.182 & $3.07 \pm 0.09$ &  $1.63\pm0.12$ & $\Gamma=2.51\pm0.05$ with excess absorption \\
RGB J0710+591    & 0.125 & $2.69 \pm 0.26 \pm 0.20$ & $1.46\pm0.22$ & $\Gamma=1.86\pm0.01$ with Galactic absorption \\
1ES 0414+009       & 0.287 & $3.5 \pm 0.3\pm 0.2$ &  $1.9 \pm 0.1$ & $\Gamma=2.4\pm0.1$ with excess absorption \\
H 2356$-$309         & 0.165 & $3.06 \pm 0.15 \pm 0.10$ & $1.89\pm0.17$ &  $\Gamma_{low}=2.08\pm0.03$, $\Gamma_{high}=2.32\pm0.02$, $E_{brk}=1.00\pm0.08$\,keV \\
\hline
\end{tabular}
\end{center}
\tablecomments{$^{\dagger}$ Spectral information in lowest X-ray flux is taken from literature.}
\tablecomments{Spectral information for the lowest X-ray flux is taken from the literature. References: HESS J1943+213 \citep{HESS2011}, and this work. 1ES~0229+200 \citep{HESS2007, Vovk12, Kaufmann11}. 1ES~0347$-$121 \citep{NewHESS, Perlman05}, and this work. 1ES~1101$-$232 \citep{HESS1101, Ackermann11, Reimer08}. 1ES~1218+304 \citep{VERITAS1218, Donato05}. RGB~J0710+591 \citep{VERITAS0710}. 1ES~0414+009 \citep{VERITAS0414}. H~2356$-$309 \citep{HESS2356, 2LAC}.
}
\end{table*}

\subsection{Spectral Modeling}
\label{sec-IGMF}

If \hess\ is indeed an extreme HBL, its unique properties --- in particular its steady hard X-ray continuum together with very soft VHE spectrum --- may be relevant in the context of constraining the IGMF intensity \citep{Neronov10,Tavecchio11,Dermer11}. Unfortunately, the intense diffuse Galactic emission in the direction of this source hampers its high significance detection in the \fermi\ range, which is crucial for such an analysis. Extreme HBLs located at higher Galactic latitudes, such as \newhs, could therefore be more suitable for this purpose. Hence, we first report on the spectral modeling of \newhs\ including the new \fermi\ detection reported in this paper. We then repeat the same modeling for \hess\ assuming its blazar nature, predominantly in order to access the source energetics.

In the fitting procedure we adopt a model that includes the radiation mechanisms of synchrotron and synchrotron self-Compton \citep[SSC; for details of the model and the $\chi^2$ minimization technique see][]{Finke08}. Additionally, we include a cascade component, from inverse-Compton upscattering of cosmic microwave background (CMB) photons by the electron-positron pairs created by primary TeV photons absorbed by the EBL. This component was calculated using the method described by \citet{Dermer11}, with the \citet{Finke10_EBL} EBL model, a jet opening angle of 0.1 rad, and the IGMF coherence length of $\lambda_B = 1$\,Mpc, assuming there is no significant amount of $\gamma$-rays from ultra-high energy cosmic rays interacting with the CMB and EBL \citep[e.g.,][]{Essey10}, and no significant synchrotron energy losses for the created electron-positron pairs \citep{Broderick12}. We assume that both \newhs\ and \hess\ have been emitting VHE $\gamma$-rays at their current levels for 3\,yr and $\sim$\,1 M yr, respectively, from the jet regions characterized by the light-crossing timescale of $10^5$\,s.

\begin{figure}
\begin{center}
\plotone{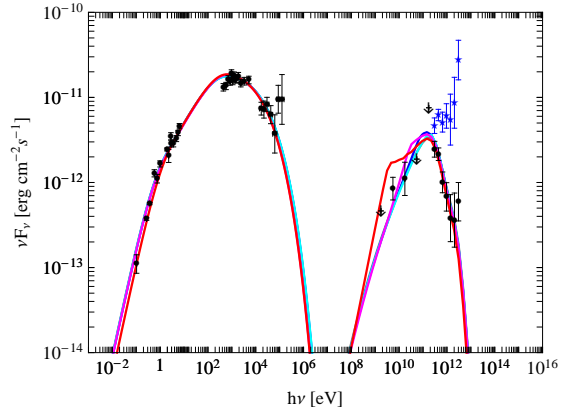}
\caption{Broadband SED of \newhs. Black filled circles/arrows correspond to the same data as shown in Figure~\ref{fig:SED2} (see also \S~\ref{newhs-data} and \ref{sec-newGeV}). Blue stars denote the H.E.S.S. fluxes of the source corrected for the EBL absorption using the model of \citet{Finke10_EBL}. The model curves included in the figure represent the sum of all the considered emission components (synchrotron, SSC, and cascade), for different values of the IGMF, namely $B_{\rm IG} = 10^{-16}$\,G (gray), $B_{\rm IG} = 10^{-17.5}$\,G (blue), $B_{\rm IG} = 10^{-17}$\,G (cyan), $B_{\rm IG} = 10^{-18}$\,G (magenta), and $B_{\rm IG} = 10^{-19}$\,G (red; see \S~\ref{sec-IGMF} for the model fit description). \label{fig:0347SED}}
\end{center}
\end{figure}

The resulting model curves of \newhs\ for different values of the IGMF between $B_{\rm IG} = 10^{-16}$\,G and $10^{-19}$\,G are shown in Figure~\ref{fig:0347SED}. The corresponding model parameters are listed in Table~\ref{table_0347}. Note that the comoving size of the emission blob $R^{\prime}_b$ is calculated from a given variability timescale $t_{var}$ (model parameter) as $R^{\prime}_b=c \delta t_{var}/(1+z)$, where $c$ is the speed of light, $\delta$ is the beaming factor of the jet, and $z$ is the redshift of the source. Also, the electron distribution is assumed to be of a broken power-law form, with low- and high-energy photon indices denoted as $s_1$ and $s_2$, respectively. The best fit, as determined using the reduced $\chi^2$ statistic, is formally provided by the model with $B_{\rm IG}=10^{-16}$\,G, for which the cascade is minimized. Any higher $B_{\rm IG}$ would give an identical fit due to a negligible cascade contribution. Given the model and all the caveats and assumptions that go into it, it seems therefore likely that $B_{\rm IG} \ga 3 \times 10^{-17}$\,G. If $\lambda_B$ is lower than the assumed 1\,Mpc, the resulting constraint on the IGMF would be stronger, since $B_{\rm IG} \propto \lambda_B^{-1/2}$. If, however, $\lambda_B > 1$\,Mpc, the constraint would be essentially unchanged, since in this case the electrons lose basically all of their energy within one correlation length \citep[e.g.,][]{Neronov09}. Our lower limit is about two orders of magnitude below that derived by \citet{Tavecchio11}. The reason for this is that \citeauthor{Tavecchio11} assumed that the $\gamma$-ray emission of \newhs\ is stable over $10^7$ year, while in our case 3 yr is assumed. We note that, if the engine timescale is longer (say 10 or 1000 yr) instead of 3 yr as we assumed here, there will be more cascade emission, and the IGMF has to be larger to keep the cascade below the LAT emission for \newhs. Therefore, the assumption of a short engine timescale is more conservative, since it provides a weaker constraint on the IGMF. Future high-sensitivity observations in the TeV range using Cherenkov Telescope Array, for example, would provide much tighter constrains on this crucial variability timescale. On the other hand, extremely low \emph{jet} magnetic field and high minimum electron Lorentz factor, both of which were claimed for the source by \citet{Tavecchio11}, are confirmed in our modeling (see Table~\ref{table_0347}).

\begin{figure} 
\begin{center}
\plotone{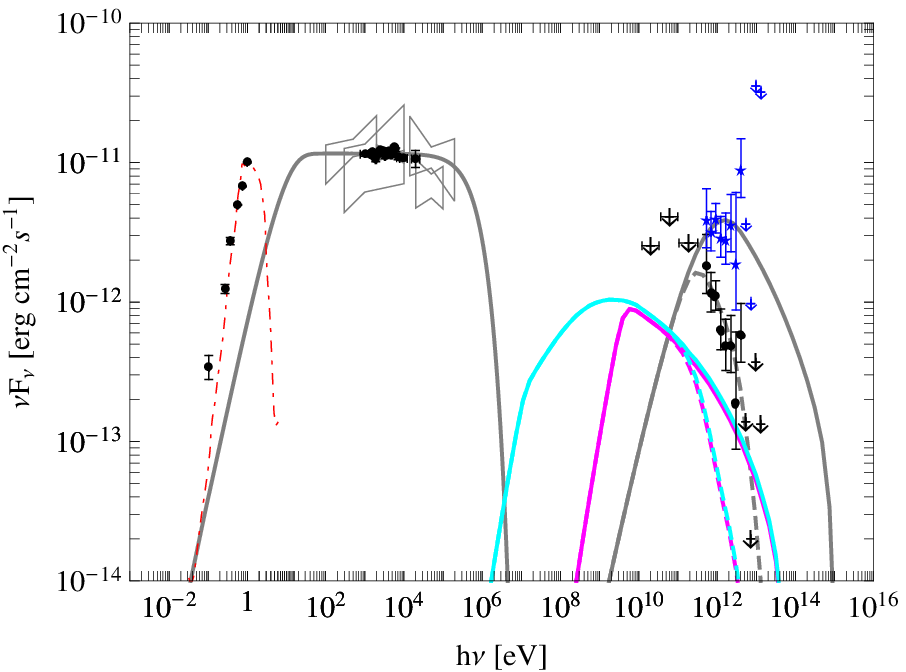}
\caption{Broadband SED of \hess. Black filled circles/large arrows correspond to the same data as shown in Figure~\ref{fig:SED2} (see also \S~\ref{hs-data} and \ref{sec-GeV}). Blue stars and small arrows denote the de-absorbed VHE spectrum, assuming a source distance of $600$\,Mpc and the EBL model of \citet{Finke10_EBL}. Various model curves represent different emission components calculated in the framework of the `extreme HBL fit' (see \S~\ref{sec-IGMF}), including synchrotron and SSC ones (gray curves), along with the cascade spectra calculated for $B_{\rm IG} = 10^{-16}$\,G (magenta curves) and $B_{\rm IG} = 10^{-18}$\,G (cyan curves); the dashed curves denote the spectra with the EBL attenuation included; thin red dot-dashed curve illustrates the starlight of a giant elliptical galaxy located at a distance of 600\,Mpc. \label{fig:1943SED}}
\end{center}
\end{figure}

In the modeling of \hess\ (see Figure~\ref{fig:1943SED}), we assumed the electron energy distribution in a single power-law form with index $s=3$ between electron Lorentz factors $\gamma_{min} = 10^5$ and $\gamma_{max} = 3 \times 10^7$ (in agreement with the \texttt{SYNCHROTRON} modeling presented in \S~\ref{sec-X}). The model parameters are tabulated in Table~\ref{table_1943}. The fit --- characterized by the reduced $\chi^2$/dof\,$= 5.8/6$ based on the SSC match to the H.E.S.S. data only --- returns the jet Doppler factor $\delta = 70$ (the maximum value from the assumed range of this model parameter) and the emission region magnetic field $B = 0.78$\,mG, which are both consistent with the analogous values derived for the other extreme HBLs \citep{Tav10}. The corresponding total jet kinetic luminosity carried by the radiating electrons is $P_{j,\,e} \simeq 6.3 \times 10^{44}$\,erg\,s$^{-1}$. The calculated cascade spectra are as expected considerably below the newly derived \fermi\ upper limits (as long as the IGM is not orders of magnitudes lower than $10^{-18}$\,G).

\section{Conclusion}
\label{sec-concl}

Based on the new \suz\ and \fermi\ data, augmented by the infrared photometry from the \wise\ and \uki\ surveys, here we argue that the `extreme HBL' scenario for \hess\ remains the most plausible option. Our conclusion follows from (i) the derived best-quality X-ray spectrum of the source revealing excess absorption, (ii) the fact that the combined \wise\ and \uki\ spectrum of the infrared counterpart of the source is consistent with a luminous elliptical host located at the distance of $\sim 600$\,Mpc, (iii) the broad-band spectral fitting returning the model parameters in agreement with the analogous ones claimed for the established extreme HBLs, and also (iv) a close resemblance between the broad-band SEDs of \hess\ and other extreme HBLs, in particular \newhs. In addition, lack of any GeV counterpart of \hess\ in the \fermi\ data, but on the other hand a presence of arcmin-scale radio halo, can then be easily understood as well. Yet the persistent hard X-ray emission together with particularly low brightness temperature of the radio core seem to be at odds, at least to some extent, with the blazar nature of the source.

Here we also report on the detection of \newhs\ in the most recent accumulation of the \fermi\ data, noting that this object was the only known extreme HBL missing thus far its GeV counterpart. The newly derived GeV spectrum of the source, along with the updated hard X-ray and infrared fluxes, enabled detailed broad-band modeling including the cascade component due to the electron-positron pairs created by the EBL-absorbed TeV jet emission. We confirm the extremely low jet magnetic field and high minimum electron Lorentz factor reported by \citet{Tavecchio11}. However, our model fits yield more conservative lower limits for the IGMF intensity ($\ga 3 \times 10^{-17}$\,G) than those derived previously by \citet{Tavecchio11}.

\begin{table*}
\footnotesize
\begin{center}
\caption{Model parameters for the SED of \newhs\ shown in Figure~\ref{fig:0347SED}.
\label{table_0347}}
\begin{tabular}{lcccccc}
\hline
Parameter & Symbol & \multicolumn{5}{c}{Values}\\
\hline
\hline
Intergalactic Magnetic Field [G] & $B_{\rm IG}$ & $10^{-19}$ & $10^{-18}$ & $3\times10^{-18}$ & $10^{-17}$ & $10^{-16}$ \\
\hline
Reduced $\chi^2$ & $\chi^2$/dof  & 17/7 & 9.6/7 & 10/7 & 11/7 & 9.5/7 \\
\hline
Bulk Lorentz Factor & $\Gamma$	 & 49 & 62 & 61 & 62 & 61  \\
Doppler Factor & $\delta_D$	 & 49 & 62 & 61 & 62 & 61  \\
Blazar Magnetic Field [mG] & $B$ & 3.1  & 1.3 & 1.4 & 1.3 & 1.3   \\
Variability Timescale [s]& $t_v$       & 1$\times$$10^5$ & 1$\times$$10^5$ & 1$\times$$10^5$ & 1$\times$$10^5$ & 1$\times$$10^5$  \\
Comoving radius of blob [cm]& R$^{\prime}_b$ & 1.2$\times$10$^{17}$ & $1.6\times10^{17}$ & $1.5\times10^{17}$ & $1.6\times10^{17}$ & $1.6\times10^{17}$ \\
\hline
Low-Energy Electron Spectral Index & $s_1$       & 2.0 & 2.0 & 2.0 & 2.0 & 2.0     \\
High-Energy Electron Spectral Index  & $s_2$       & 2.8 & 2.8 & 2.8 & 2.8 & 2.8   \\
Minimum Electron Lorentz Factor & $\gamma^{\prime}_{min}$  & $2.0\times10^4$ & $2.0\times10^4$ & $2.0\times10^4$ & $2.0\times10^4$ & $2.0\times10^4$  \\
Break Electron Lorentz Factor & $\gamma^{\prime}_{brk}$ & $4.3\times10^5$ & $6.0\times10^5$ & $5.2\times10^5$ & $5.2\times10^5$ & $5.4\times10^5$ \\
Maximum Electron Lorentz Factor & $\gamma^{\prime}_{max}$  & $2.3\times10^6$ & $3.1\times10^6$ & $3.4\times10^6$ & $3.5\times10^6$ & $3.5\times10^6$ \\
\hline
Jet Power in Magnetic Field [erg s$^{-1}$] & $P_{j,B}$ & $2.6\times10^{42}$ & $1.3\times10^{42}$ & $1.3\times10^{42}$ & $1.3\times10^{42}$ & $1.2\times10^{42}$  \\
Jet Power in Electrons [erg s$^{-1}$] & $P_{j,e}$ & $5.1\times10^{44}$ & $1.0\times10^{45}$ & $1.0\times10^{45}$ & $1.1\times10^{45}$ & $1.1\times10^{45}$ \\
\hline
\end{tabular}
\end{center}
\end{table*}

\begin{table*}
\footnotesize
\begin{center}
\caption{Model parameters for the SED of \hess\ shown in Figure~\ref{fig:1943SED}.
\label{table_1943}}
\begin{tabular}{lccc}
\hline
Parameter & Symbol & \multicolumn{2}{c}{Values}\\
\hline
\hline
Intergalactic Magnetic Field [G] & $B_{\rm IG}$ & $10^{-18}$ & $10^{-16}$ \\
\hline
Reduced $\chi^2$ & $\chi^2$/dof  & 5.7/7 & 5.7/7 \\
\hline
Bulk Lorentz Factor & $\Gamma$	 & 70 & 70  \\
Doppler Factor & $\delta_D$	 & 70 & 70  \\
Blazar Magnetic Field [mG] & $B$ & 0.78 & 0.78   \\
Variability Timescale [s]& $t_v$       & 1$\times$$10^5$ & 1$\times$$10^5$ \\
Comoving radius of blob [cm]& R$^{\prime}_b$ & 1.9$\times$10$^{17}$ & $1.9\times10^{17}$ \\
\hline
Low-Energy Electron Spectral Index & $p_1$       & 3.0 & 3.0 \\
Minimum Electron Lorentz Factor & $\gamma^{\prime}_{min}$  & $1.0\times10^5$ & $1.0\times10^5$ \\
Maximum Electron Lorentz Factor & $\gamma^{\prime}_{max}$  & $3.0\times10^7$ & $3.0\times10^7$ \\
\hline
Jet Power in Magnetic Field [erg s$^{-1}$] & $P_{j,B}$ & $7.8\times10^{41}$ & $7.8\times10^{41}$  \\
Jet Power in Electrons [erg s$^{-1}$] & $P_{j,e}$ & $6.3\times10^{44}$ & $6.3\times10^{44}$ \\
\hline
\end{tabular}
\end{center}
\end{table*}

\acknowledgments
Y.T.T is supported by Kakenhi 24840031. \L .S. was supported by Polish NSC grant DEC-2012/04/A/ST9/00083. Work by C.C.C. at NRL is supported in part by NASA DPR S-15633-Y.

The \textit{Fermi} LAT Collaboration acknowledges generous ongoing support
from a number of agencies and institutes that have supported both the
development and the operation of the LAT as well as scientific data analysis.
These include the National Aeronautics and Space Administration and the
Department of Energy in the United States, the Commissariat \`a l'Energie Atomique
and the Centre National de la Recherche Scientifique/Institut National de Physique
Nucl\'eaire et de Physique des Particules in France, the Agenzia Spaziale Italiana
and the Istituto Nazionale di Fisica Nucleare in Italy, the Ministry of Education,
Culture, Sports, Science and Technology (MEXT), High Energy Accelerator Research
Organization (KEK) and Japan Aerospace Exploration Agency (JAXA) in Japan, and
the K.~A.~Wallenberg Foundation, the Swedish Research Council and the
Swedish National Space Board in Sweden.

Additional support for science analysis during the operations phase is gratefully
acknowledged from the Istituto Nazionale di Astrofisica in Italy and the Centre National d'\'Etudes Spatiales in France.

\bibliographystyle{apj}

\end{document}